\documentclass[a4paper,10pt,twoside]{article}

\usepackage[english]{babel}
\usepackage{amsmath}
\usepackage{amssymb}
\usepackage{epsfig}
\usepackage{mathrsfs}
\usepackage{graphicx}
\DeclareMathOperator{\asin}{asin}
\usepackage{amssymb}
\usepackage{subfigure}
\usepackage{multirow}
\usepackage{float} 
\usepackage{ccaption} 
\usepackage[top=2cm,left=1.5cm,bottom=2cm,right=1.5cm,columnsep=23pt]{geometry}

\usepackage{multicol} 
\usepackage{float} 
\usepackage{hyperref} 

\usepackage{abstract} 

\usepackage{titlesec} 

\usepackage{fancyhdr} 

\date{}

\begin{document}
\pagestyle{empty}
\begin{multicols}{2}

\noindent \textbf{Three-dimensional orientation measurement of a single fluorescent nanoemitter by polarization analysis}\\

\noindent Clotilde Lethiec$^{1,2}$, Julien Laverdant$^{1,2,3}$, Henri Vallon$^{1,2}$, Cl\'ementine Javaux$^{4}$, Beno\^it Dubertret$^{4}$, Jean-Marc Frigerio$^{1,2}$, Catherine Schwob$^{1,2}$, Laurent Coolen$^{1,2}$, Agn\`es Ma\^itre$^{1,2}$\\

\noindent $^1$ Universit\'e Pierre et Marie Curie-Paris 6, UMR 7588, INSP, 4 place Jussieu, Paris cedex 05, France \\ $^2$ CNRS, UMR 7588, INSP, 4 place Jussieu, Paris cedex 05, France \\ $^3$ Laboratoire PMCN, Universit\'e de Lyon, Universit\'e Lyon-1, UMR CNRS 5586, 69622 Villeurbanne, France \\ $^4$ LPEM, ESPCI/CNRS/UPMC UMR 8213, 10, rue Vauquelin, 75005 Paris, France.

\begin{figure}[H]
\begin{center}\includegraphics[width=6cm]{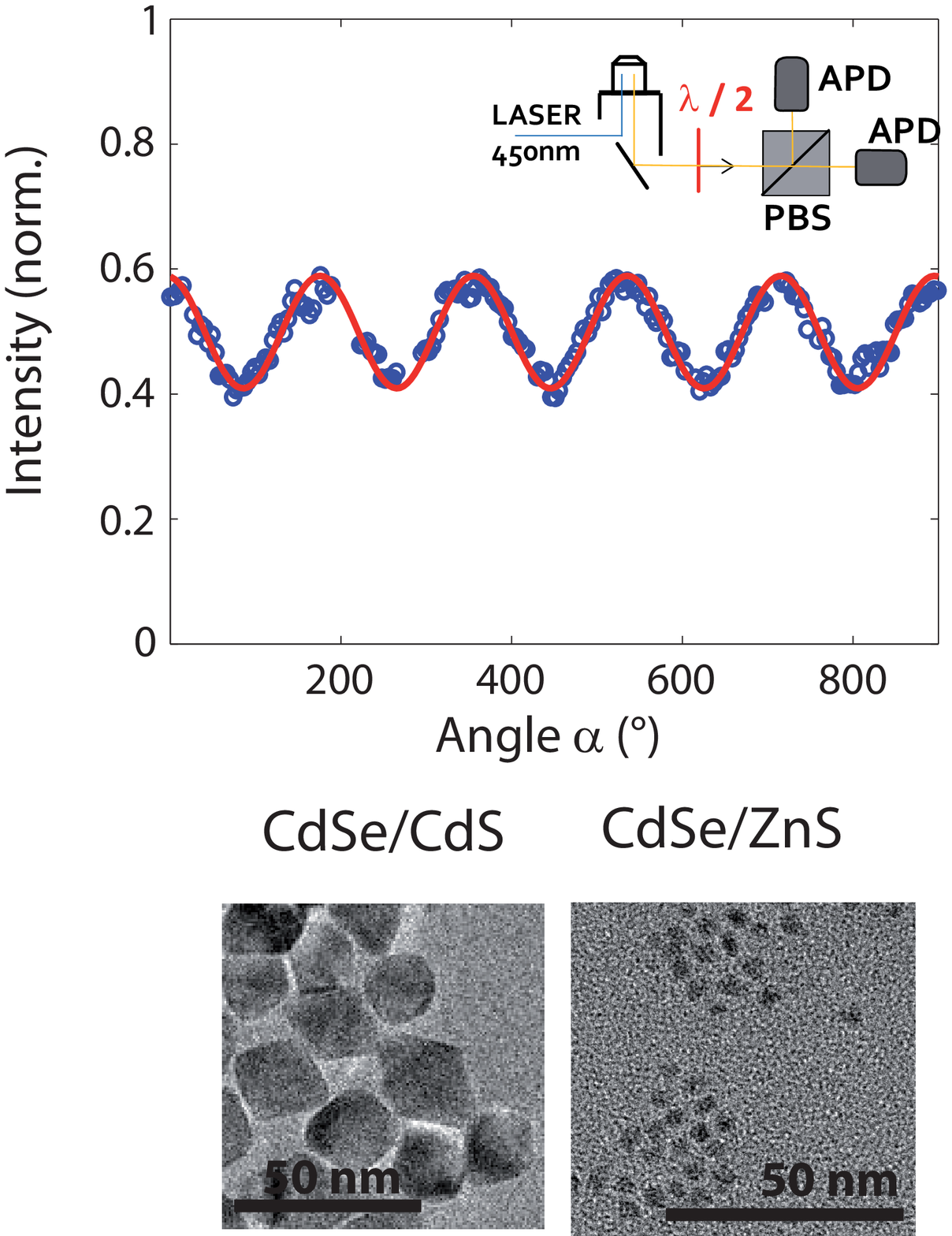}\end{center}
\legend{}
\end{figure}

The orientation of a single dipole is a crucial data in various fields, notably in biology and nanophotonics. We show that it is possible to retrieve the 3D-orientation of a single dipole by using a polarimetry method. We develop a formalism which takes into account the nature and environment of the nanoemitter and perform experimental analysis on high-quality colloidal CdSe/CdS nanocrystals.

\end{multicols}

\title{\vspace{-15mm}\fontsize{24pt}{10pt}\selectfont\textbf{Three-dimensional orientation measurement of a single fluorescent nanoemitter by polarization analysis}}

\author{
Clotilde Lethiec$^{1,2}$, Julien Laverdant$^{1,2,3}$, Henri Vallon$^{1,2}$, Cl\'ementine Javaux$^{4}$,\\ Beno\^it Dubertret$^{4}$, Jean-Marc Frigerio$^{1,2}$, Catherine Schwob$^{1,2}$, Laurent Coolen$^{1,2}$\\ and Agn\`es Ma\^itre$^{1,2}$  \\
\\ $^1$Universit\'e Pierre et Marie Curie-Paris 6, UMR 7588, INSP, 4 place Jussieu,\\ Paris cedex 05, France \\ $^2$ CNRS, UMR 7588, INSP, 4 place Jussieu,\\ Paris cedex 05, France \\ $^3$ Laboratoire PMCN, Universit\'e de Lyon, Universit\'e Lyon-1, UMR CNRS 5586,\\ 69622 Villeurbanne, France \\ $^4$ LPEM, ESPCI/CNRS/UPMC UMR 8213, 10, rue Vauquelin, 75005 Paris, France.\\
\\ \normalsize \href{agnes.maitre@insp.jussieu.fr}{Address correspondence to  agnes.maitre@insp.jussieu.fr}	
}

\maketitle
\thispagestyle{fancy}

\begin{abstract}
We demonstrate theoretically and experimentally that the three-dimensional orientation of a single fluorescent nano-emitter can be determined by polarization analysis of the emitted light (while  excitation polarization analysis provides only the in-plane orientation). The determination of the emitter orientation by polarimetry requires a theoretical description including the objective numerical aperture, the 1D or 2D nature of the emitting dipole and the environment close to the dipole. We develop a model covering most experimentally relevant microscopy configurations and provide analytical relations useful for orientation measurements. We perform polarimetric measurements on high-quality core-shell CdSe/CdS nanocrystals and demonstrate that they can be approximated by two orthogonal degenerated dipoles. Finally, we show that the orientation of a dipole can be inferred by polarimetric measurement even for a dipole in the vicinity of a gold film, while in this case the well-established defocused microscopy is not appropriate.
\end{abstract}


\begin{multicols}{2} 

The determination of the orientation of a single photoluminescent emitter is a major issue since early single-molecule studies \cite{Ha96,Nie,Trautman,Betzig93,Orrit}. It is a valuable tool for understanding distortion mechanisms in polymers \cite{Trabesinger00} or biological systems \cite{Forkey03, Toprak06}. For nano-optics and plasmonics, the orientation of an emitter has a strong influence on its coupling to the environment \cite{Brokmann05, Macklin96, Vion09,Bharadwaj,Belacel12}.

First orientation studies relied on polarised excitation \cite{Ha99}. The electric field at the focal point thus probes mainly the in-plane component of the dipole, and its azimuthal orientation $\Phi$ can be inferred, but the (out-of-plane) polar orientation $\Theta$ remains unknown \cite{Guttler96}. In order to measure $\Theta$, various sophisticated schemes have been proposed to increase the out-of-plane component of the electric field \cite{Betzig93, Lieb01, Novotny01, Sick00, Dickson98, Ishitobi}.   

Moreover, most of these works intrinsically probe the orientation of the absorbing dipole, which for non-resonant photoluminescence can be extremely different from the emitting dipole. Depending on their group symmetry, molecules can have orthogonal excited and emitting dipoles \cite{Pariser56}. As for colloidal semiconductor nanocrystals, while they show little dependence on the excitation polarization \cite{Chizhik11}, because of their dense continuum of absorption levels, their emission is usually described as a sum of two incoherent orthogonal dipoles (referred to as "two-dimensional (2D) dipole") \cite{Empedocles99, Chung03, Brokmann05b}. The orientation of such a nanocrystal can thus only be obtained from its emission properties, as its excitation properties are isotropic. Various methods have been suggested to determine the orientation $\Theta$ of emitting dipoles. They rely mostly (except \cite{Macklin96} which uses relative lifetime differences) on probing the emission diagram by decomposing the different emission directions with an annular separating plate \cite{Hohlbein08} or taking advantage of standard \cite{Dickson98} or tailored aberrations \cite{Bartko99b}. In the latter category, a successful method has been defocused imaging \cite{Toprak06, Brokmann05,Jasny97, Bartko99, Bohmer03, Sepiol97, Patra04,Lieb04}, which offers the most convenient implementation. Recently, several authors have extended this method to super-resolution experiments \cite{Aguet,Backlund}. Possibly by analogy with the idea that the out-of-plane component of the \textit{excitation} dipole cannot be probed by polarization analysis, for the \textit{emission} dipole just few studies have considered polarization analysis for orientation measurements \cite{Empedocles99,Fourkas01}.

In this paper, we show that the three-dimensional orientation $(\Theta, \Phi)$ of a nano-emitter can be obtained by analysing its emission polarization, for both 1D and 2D dipoles, provided that the objective numerical aperture is sufficient and that a theoretical analysis is performed. We insist on the fact that the emitter optical environment (such as proximity to an interface) must be taken into account. We provide an analytical model that can be used to interpret the data and extract $(\Theta, \Phi)$ in a wide range of realistic experimental conditions. We demonstrate this experimentally by measuring the orientation of high-quality thick-shell CdSe/CdS nanocrystals with 2D-dipolar emission, including the case when the emitter lies in proximity to a gold film, a situation for which the more standard defocused imaging is not sufficiently sensitive to provide reliable information. 

\pagestyle{plain}
In the first section, we present the elements of our theoretical model. In a second section, we detail the difference between 1D and 2D dipoles and show experimentally that our CdSe/CdS nanocrystals are 2D dipoles. In the third section, we develop the results of the theoretical model and show that the orientation can be extracted from polarization data. We implement this experimentally in the fourth section and measure the orientation of various nanocrystals. In the last section, we consider the case of a nanocrystal near a metallic film, and show that the orientation can be obtained from polarization analysis but not from defocused imaging.


\section{Theoretical framework}

\begin{figure}[H]
\begin{center}\includegraphics[width=6cm]{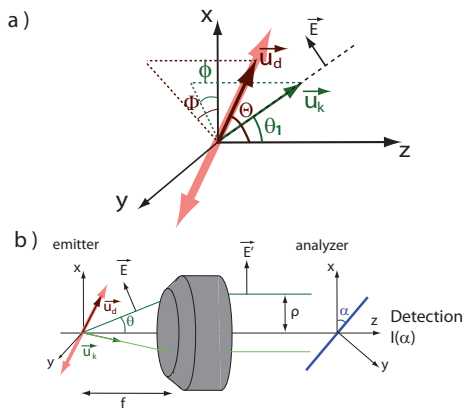}\end{center}
\caption{a) Schematic of a normalized dipole $\vec{u}_d$ with in-plane angle $\Phi$ and off-axis angle $\Theta$ and a normalized wave vector $\vec{u}_k$  with in-plane angle $\phi$ and off-axis angle $\theta_1$. b)Schematic of the simulated situation: the dipole described above, a microscope objective, and a polarizer with in-plane angle $\alpha$.}
\label{Fig1}
\end{figure}

In this section we develop the theoretical modelling used in this study. The simulated situation is illustrated on Fig.\ref{Fig1}(a) and (b)\,: the emission of a dipole is collected by an objective and analysed by a rotating polarizer. The polarizer orientation angle $\alpha$ is continuously rotated. The principle of the measurement is to extract, from the detected intensity $I(\alpha)$, the azimuthal angle $\Phi$ and the polar angle $\Theta$ of the emitting dipole. 

The far-field emission component of the electric field emitted by a linear dipole (1D-dipole) into a direction $(\theta_1, \phi)$ can be expressed as\,:

\begin{equation}
\vec{E}(\theta_1, \phi) = \frac{D}{r}(\vec{u}_k \wedge \vec{u}_d \wedge \vec{u}_k)
\label{eq1}
\end{equation}

with $D$ a constant depending on the refractive index of the medium containing the dipole, where\,:
\begin{align}
\vec{u}_d &= \left(\begin{matrix}\sin\Theta\cos\Phi \\ \sin\Theta\sin\Phi \\ \cos\Theta \end{matrix}\right)\
\\ \vec{u}_k &= \frac{1}{k_1} \left(\begin{matrix} k_{1_\parallel}\cos\phi\\ k_{1_\parallel}\sin\phi \\ k_{1_z}\end{matrix}\right) = \left(\begin{matrix}\sin\theta_1\cos\phi\\ \sin\theta_1\sin\phi \\ \cos\theta_1\end{matrix}\right)\label{eq2}
\end{align}
are the unit vectors corresponding respectively to the dipole orientation and to the considered emission $\vec{k_1}$-vector direction.\\
Equation (\ref{eq1}) expresses the emission of a dipole in a homogeneous dielectric environment, which can be decomposed into its s and p components\,:
\begin{equation}
\vec{E}(\theta_1, \phi) = \frac{D}{r} (E_s(\theta_1,\phi)\vec{u}_s + E_p(\theta_1,\phi)\vec{u}_p)
\label{eq3}
\end{equation}
with the unit vectors\,:
\begin{equation}
\vec{u}_s = \left(\begin{matrix}\sin\phi \\ -\cos\phi \\ 0 \end{matrix}\right)\,\qquad \text{and} \qquad\,\vec{u}_p = \left(\begin{matrix}\cos\theta_1\cos\phi\\ \cos\theta_1\sin\phi \\ -\sin\theta_1\end{matrix}\right)
\end{equation}
and
\begin{align}
 E_s(\theta_1,\phi) &= \sin\Theta\sin(\phi - \Phi) \label{EqEs}
\\E_p(\theta_1,\phi) &= (E_{p_a}(\theta_1,\phi) + E_{p_b}(\theta_1,\phi))
\end{align}
with\,:
\begin{align}
E_{p_a}(\theta_1,\phi) &= - \cos\Theta\sin\theta_1 \label{EqEpa}
\\E_{p_b}(\theta_1,\phi) &= \sin\Theta\cos\theta_1\cos(\Phi - \phi) \label{EqEpb}
\end{align}

\begin{figure}[H]
\begin{center}\includegraphics[width=6cm]{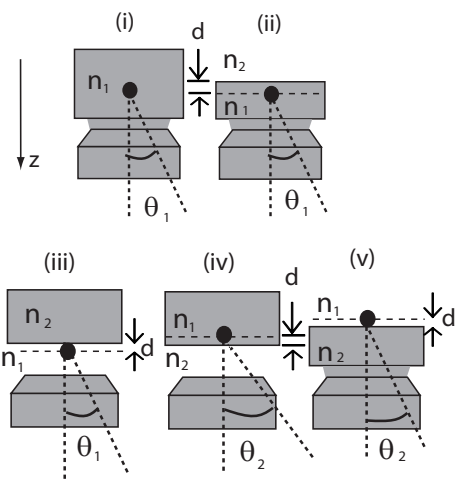}\end{center}
\caption{The five cases corresponding to different experimental conditions, numbered from (i) to (v).}
\label{Fig1bis}
\end{figure}

However, in many experimental observation conditions, the dipole is in the vicinity of an optical interface, which modifies the emission diagram and polarization \cite{Lukosz81}. We describe here five relevant experimental conditions, as presented in Fig.\ref{Fig1bis}. Aside from the case of a homogeneous medium (i), we consider\,:

\begin{itemize}
\item (ii) a sample with emitters deposited on a planar substrate of index $n_1$ and protected by a polymer layer of thickness $d$ of same index, observed with an immersion objective, the upper medium (most likely air) being of index $n_2$,
\item (iii) a sample with emitters at a distance d (with d tending towards 0) from a planar surface (substrate index $n_2$) without any protecting layer, observed with an air objective (air index $n_1$),
\item (iv) a sample with emitters on a planar surface with a polymer protecting layer (index $n_1$), observed with an air objective (air index $n_2$),
\item (v) a sample with emitters at a distance d (with d tending towards 0) from a planar surface (index $n_1$) without any protecting layer, observed with an immersion objective (index $n_2$).
\end{itemize}

For all situations, we write $n_1$ the index of the medium containing the emitter (glass index for (i), (ii), (iv) and air index for (iii) and (v)) and $n_2$ the index of the other medium (glass index for (iii) and (v) and air index for (ii) and (iv)).\\

For situations (ii) and (iii), the detected electric field is a sum of the direct emitted and its reflected fields, whereas in situations (iv) and (v) only the direct emission is collected, after transmission through the interface. In the last situation (v), we need to take into account for the detected field the evanescent component of the dipole's near-field emission, which becomes propagative when transmitted through the interface \cite{Novotny}. This particular case is developed in the Appendix. \\

The presence of a near interface can be described generally by multiplying $E_s$, $E_{p_a}$ and $E_{p_b}$ by the functions $f_s$, $f_{p_a}$ and $f_{p_b}$, whose definition depends whether it involves reflection or transmission of the emitted field \cite{Lukosz81}\,:

\begin{center}
\begin{tabular}{|c|c|c|c|}
\hline
   & $f_s =$ & $f_{p_a} =$ & $f_{p_b} =$  \\ \hline 
(i) & 1 & 1 & 1 \\
(ii) & $1 + r_s^{12}e^{i\Delta}$ & $1 + r_p^{12}e^{i\Delta}$ & $1 - r_p^{12}e^{i\Delta}$\\
(iii) & $1 + r_s^{12}e^{i\Delta}$ & $1 + r_p^{12}e^{i\Delta}$ & $1 - 
r_p^{12}e^{i\Delta}$ \\
(iv) & $t_s^{12}$ & $t_p^{12}$ & $t_p^{12}$ \\
\hline
\end{tabular}\end{center}

with $\Delta(\theta_1) = 4\pi n_1d\cos(\theta_1)/\lambda$ and the Fresnel reflection and transmission coefficients\,:

\begin{align}
  r_s^{12} &= \frac{n_1\cos\theta_1 - n_2\cos\theta_2}{n_1\cos\theta_1 + n_2\cos\theta_2}
\\  r_p^{12} &= \frac{n_2\cos\theta_1 - n_1\cos\theta_2}{n_1\cos\theta_2 + n_2\cos\theta_1}
\\  t_s^{12} &= \frac{2n_1\cos\theta_1}{n_1\cos\theta_1 + n_2\cos\theta_2}
\\  t_p^{12} &= \frac{2n_1\cos\theta_1}{n_1\cos\theta_2 + n_2\cos\theta_1}
\end{align}

with 
\begin{equation}
n_1\sin\theta_1 = n_2\sin\theta_2
\end{equation}

\bigskip
The field emitted by the point-like dipole collected by the objective is collimated after the lens.
This field becomes after passage through the objective of focal length f (neglecting aberrations) and taking into account the apodization factor $\left(\cos\theta_{j}\right)^{-\frac{1}{2}}$ \cite{Novotny}, where $j=1$ in the cases (ii) and (iii) and $j=2$ in cases (iv) and (v)\,:

\begin{equation}
\begin{split}
{\vec{E'}}(\theta_{j}, \phi) = \frac{1}{f}\left(\cos\theta_{j}\right)^{-\frac{1}{2}} \vec{\mathcal{E}}
\end{split}
\end{equation}

where\,:

\begin{equation}
\begin{split}
\vec{\mathcal{E}}(\theta_1,\phi) = & D \frac{n_j}{n_1} (E_s(\theta_1,\phi)f_s(\theta_1)\vec{v}_s \\ & + [E_{p_a}(\theta_1,\phi)f_{p_a}(\theta_1)+E_{p_b}(\theta_1,\phi)f_{p_b}(\theta_1)]\vec{v}_p)
\label{eqErond}
\end{split}
\end{equation}

with the new unit vectors\,:

\begin{equation}
\vec{v}_s = \vec{u}_s \qquad \text{and} \qquad \vec{v}_p = \left(\begin{matrix} \cos\phi \\ \sin\phi \\ 0 \end{matrix}\right)
\end{equation}

The objective collects the beams for all values of $\phi$ between 0 and 2$\pi$ and for $\theta_{j}$ between 0 and $\theta_{jmax}$. The maximum collection angle $\theta_{jmax}$ is related to the objective numerical aperture $NA$ by $\theta_{jmax} = \asin(NA/n_{j})$.\\

\bigskip
Finally, a polarizer is set after the lens, along a unit vector $\vec{u}_{\alpha}$ at an angle $\alpha$ from the $x$-axis, so that the normalized emitted power detected after the polarizer is\,:

\begin{equation}
P(\alpha) = \int_{\rho = 0}^{\rho_{max}} \int_{\phi = 0}^{2\pi} |{\vec{{E'}}(\theta_j,\phi)}.\vec{u}_{\alpha}|^2 \rho d\rho d\phi
\label{eq13}
\end{equation}\\

with the sine condition \cite{Hecht}\,:
\begin{equation}
\rho=f\sin\theta_{j}
\end{equation}

which leads to\,: 

\begin{equation}
P(\alpha) = \int_{\theta_{j} = 0}^{\theta_{jmax}} \int_{\phi = 0}^{2\pi} f^2 |{\vec{E'}(\theta_{j},\phi)}.\vec{u}_{\alpha}|^2 \cos\theta_{j} \sin\theta_{j} d\theta_{j} d\phi
\label{eq14}
\end{equation}

For the transmission case (iv), the conservation of the power per solid angle at the interface is assured by taking into account the apodization factor $(n_1/n_2)(\cos\theta_2/\cos\theta_1)^2$.\\

Finally, for cases (i) to (iv), the emitted power is expressed as\,:

\begin{equation}
P(\alpha) = \int_{\theta_1 = 0}^{\theta_{1max}} \int_{\phi = 0}^{2\pi} |{\vec{\mathcal{E}}(\theta_1,\phi)}.\vec{u}_{\alpha}|^2 \left(\frac{n_1}{n_j}\right)^3 \frac{\cos\theta_j}{\cos\theta_1}\sin\theta_1 d\theta_1 d\phi
\label{power}
\end{equation}

For the particular case (v), the calculations are detailed in the Appendix.


\section{Determination of the emitting dipole dimension}

We have considered up to now the case of a standard linear dipole, which will be called hereafter "1D dipole". However, in many cases the emission originates from two degenerate states of orthogonal orientations. The emission is then an incoherent sum of two orthogonal 1D dipoles and is referred to as "2D dipole". Such a situation has been reported for some molecules such as benzene \cite{Pariser56}, for some nitrogen-vacancy centres \cite{Epstein05} and for CdSe/ZnS colloidal nanocrystals at low \cite{Empedocles99} and room temperature \cite{Chung03, Brokmann05b}. The latter observation has been related to theoretical calculations of the emitting state fine structure \cite{Efros96} predicting that the lowest allowed transition was twofold degenerate.

\begin{figure}[H]
\centering
\includegraphics[width=8cm]{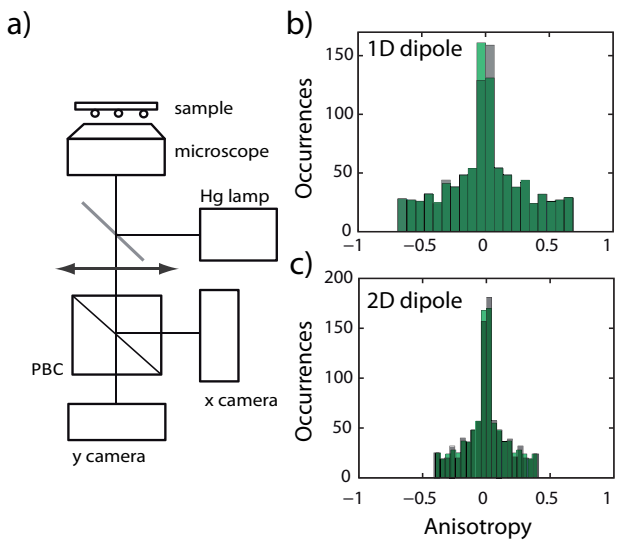} 
\caption{(a): Schematic of the set-up. A polarizing beam splitter cube (PBC) is placed in front of two CCD-cameras (directions x and y). b) and c) : Distribution of the anisotropy $\mathcal{A}= (I_x - I_y)/(I_x + I_y)$ simulated for randomly-oriented 1D-dipoles and 2D-dipoles. Calculations performed for situation (ii) with $n_1$ = 1.5, $n_2$ = 1, $d$ = 50\,nm, $\lambda$ = 620 (grey) and 565 \,nm (green) and a 1.4 numerical aperture. }
\label{Fig2}
\end{figure}

In this section, we characterize experimentally the 1D or 2D nature of the emitters considered in this paper. In order to measure the dimension of a dipole, we use the experimental set-up suggested by Chung et al. \cite{Chung03} (fig.\ref{Fig2} a)) which images simultaneously the x and y-polarized emission of the same emitters on two charge-coupled-device (CCD) cameras situated after a polarizing beam splitter. 
For practical reasons, the polarizing beam splitter cube in our experiment had to be placed not just after the objective lens but between the microscope and the CCD detector. However, we can consider the optical beam is quasi collimated in first approximation due to the high value of the lens magnification (x100), so that our theory remains valid.
The sample is prepared in configuration (ii)\,: a glass sample covered by $d$ = 50\,nm of polymer of index $n_1$ = 1.5, with air index $n_2$ = 1, and a numerical aperture equals to 1.4.

It is then possible to image many emitters and for each emitter to measure the intensities $I_x$ and $I_y$ on the two cameras and define a polarization anisotropy $\mathcal{A}$\,:

\begin{equation}
\mathcal{A}=\frac {I_x-I_y}{I_x+I_y}
\end{equation}

On the other hand, $\mathcal{A}$ can be calculated, for a given dipole with an orientation ($\Theta$,$\Phi$), from equation (\ref{eq14}) with\,:
\begin{align}
I_x&=I(\alpha=0)
\\ I_y&=I(\alpha=\frac {\pi}{2})
\end{align}
We assume a collection of emitters with random orientations isotropically distributed and plot a histogram of the calculated anisotropies, for 1D and 2D dipoles (Fig.\ref{Fig2}(b)\,; we plot for the two wavelengths 565 and 620\,nm as they will be relevant for the experiment below, although the results are very close). Both histograms display a peak at $\mathcal{A} = 0$ but 
the extension of the wings on either side of this peak is different\,: for a 1D-dipole, the wings extend up to $\pm$ 0.7 while for a 2D-dipole they extend only to $\pm$ 0.4. It is thus possible, by measuring the anisotropy of randomly oriented emitters and plotting a histogram of these anisotropies, to discriminate whether this type of emitters is 1D or 2D. Let us note that this calculation is performed for situation (ii) with the parameters indicated above\,: in a different configuration, the results would be quantitatively slightly different but the qualitative difference between 1D and 2D dipoles would remain.

\begin{figure}[H]
\centering
\includegraphics[width=8cm]{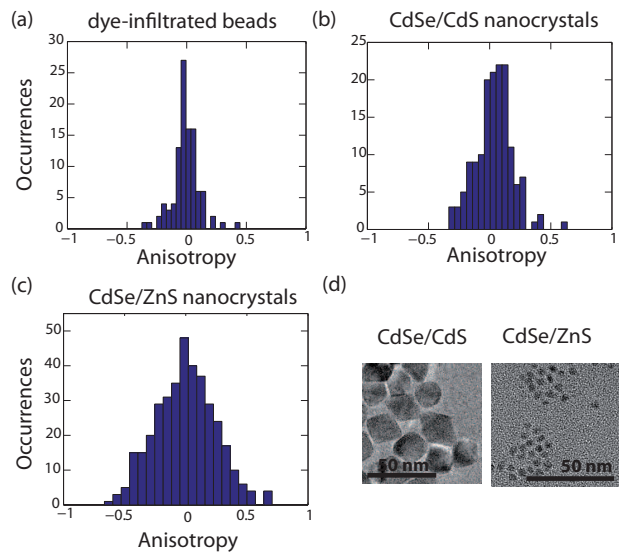}
\caption{Distribution of the anisotropy $\mathcal{A} = (I_x - I_y)/(I_x + I_y)$ measured for 103 beads (a), 152 CdSe/CdS nanocrystals (b) and 374 CdSe/ZnS nanocrystals. (c) Experimental conditions\,: situation (ii), numerical aperture 1.4, textcolor{red} {$d$ = 50\,nm} of PMMA of index 1.5. (d) Transmission electron microscopy image of the CdSe/CdS and CdSe/ZnS nanocrystal samples.}
\label{Fig3}
\end{figure}

The experimental results obtained for different types of emitters are shown on Fig.\ref{Fig3}.

Latex beads infiltrated with dye molecules (Lifetechnologies, F8763, emission peak at 600\,nm) are first studied in order to validate the method. Each bead contains a large number of emitters so that they can be considered as point like isotropic emitters: the polarization anisotropy should be zero for all beads. Indeed, we obtain (Fig.\ref{Fig3}(a), error bar: 0.05) a peak centred on $\mathcal{A} = 0$, with a width 0.05 attributed to measurement uncertainties.\\

We then study colloidal core/shell CdSe/CdS nanocrystals (emission 620nm, core diameter 2.5\,nm, total diameter 13\,nm) exhibiting very good brightness and suppressed blinking \cite{Spinicelli08} (fig. \ref{Fig3}(d)). The obtained anisotropy histogram (Fig.\ref{Fig3}(b), error bar: 0.05) presents a peak on zero and the extension of the curve reaches $\pm$0.4. This experimental curve is in satisfying agreement with the theoretical curve of Fig.\ref{Fig2}(c), taking into account a slight broadening of the central peak due to the 0.05 uncertainty on $\mathcal{A}$. This demonstrates that these nanocrystals are 2D dipoles. If they were 1D dipoles, their anisotropy histogram would extend up to $\pm$0.7.\\

Finally, we consider core/shell CdSe/ZnS nanocrystals (QDots, Invitrogen, emission 565nm) (fig. \ref{Fig3}(d)). For such nanocrystals, Empedocles et al. have reported a 2D dipole behaviour \cite{Empedocles99}. The experimental histogram obtained for these nanocrystals is plotted in Fig.\ref{Fig3}(c). The error bar is here more important (0.1) because the emission intensity is lower for these emitters. The curve presents a peak at zero and wings extending to $\pm$0.5. The agreement with the theoretical histogram for 2D dipoles is not very good, possibly because of a mixture of 1D and 2D dipolar emission as suggested in \cite{Cyphersmith10} and \cite{Early09}. This can be explained by an energy splitting,  at room temperature, smaller than $k_{B}T$ between the degenerated $\pm$1$^{L}$ and the linear 0$^{L}$ transitions, which allows a linearly polarized emission \cite{Efros96}.


\section{Theoretical results}

We now return to the main point of this paper which is the determination of the orientation of a single emitter. We elaborate here on the theory of section 1 and show that the emitting dipole orientation can be extracted from a polarization analysis. 

We distinguish between 1D and 2D dipoles as the importance of this difference was pointed out in the previous section. The orientation ($\Theta$,$\Phi$) of a 1D-dipole is defined as in section 1. For a 2D dipole, ($\Theta, \Phi$) will refer to the orientation of the emitter dark axis \cite{Empedocles99}\,: the axis normal to the plane containing the two emitting dipoles. Since the two dipoles are incoherent, the intensity emitted by a 2D dipole of dark-axis orientation ($\Theta, \Phi$) can be calculated as a sum of the intensities emitted by two 1D dipoles of orientations ($\pi/2 - \Theta, \Phi$) and ($\pi/2, \Phi + \pi/2$).
\\
 
After some manipulation, the measured intensity as a function of a rotating polarizer angle $\alpha$ (eq. \ref{eq14}) can be written as\,:
\begin{equation}
I_{1D}(\alpha) = I_{min} + (I_{max}-I_{min})\cos^2(\Phi - \alpha)
\label{eq19}
\end{equation}
\begin{equation}
I_{2D}(\alpha) = I_{max} + (I_{min}-I_{max})\cos^2(\Phi - \alpha)
\label{eq20}
\end{equation}

This expression shows that the emitted intensity is partially polarized for both 1D and 2D dipoles, with a Malus component in $\cos^2(\Phi - \alpha)$ from which the emitter in-plane orientation $\Phi$ can be straightforwardly extracted. Let us stress out however that $\Phi$ is obtained as the polarizer angle which \textit{maximizes} the detected intensity for a 1D dipole, but as the angle which \textit{minimizes} the intensity for a 2D dipole (because $\Phi$ is the orientation of the "dark axis").

We now express the maximum intensity $I_{max}$ and minimum intensity $I_{min}$ and show that their knowledge can lead to $\Theta$. For a 1D-dipole:

\begin{equation}
I_{min} = A\sin^2\Theta + B\cos^2\Theta
\label{eq21}
\end{equation}
\begin{equation}
I_{max} - I_{min} = C\sin^2\Theta
\label{eq22}
\end{equation}

and for a 2D-dipole:
\begin{equation}
I_{min} = A+B+(A-B+C)\cos^2\Theta 
\label{eq23}
\end{equation}
\begin{equation}
I_{max} - I_{min} = C\sin^2\Theta
\label{eq24}
\end{equation}

with, for both 1D and 2D dipoles, the constants\,:

\begin{equation}
A = \int_{\theta_1=0}^{\theta_{1max}} D^2\frac{\pi}{4} \frac{n_1}{n_j} \frac{\cos\theta_{j}}{\cos\theta_1} |f_s - \cos\theta_1 f_{p_b}|^2\sin\theta_1 d\theta_1
\label{eqA}
\end{equation}
\begin{equation}
B = \int_{\theta_1=0}^{\theta_{1max}} D^2\pi  \frac{n_1}{n_j}\frac{\cos\theta_{j}}{\cos\theta_1} |f_{p_a}|^2\sin^3\theta_1 d\theta_1
\label{eqB}
\end{equation}
\begin{equation}
C = \int_{\theta_1=0}^{\theta_{1max}} D^2\frac{\pi}{2} \frac{n_1}{n_j} \frac{\cos\theta_{j}}{\cos\theta_1} |\cos\theta_1 f_{p_b}+f_s|^2\sin\theta_1 d\theta_1
\label{eqC}
\end{equation}

with $j=1$ for the cases (ii) and (iii) and $j=2$ for the cases (iv) and (v).\\

From an experimental perspective, the measured $I_{max}$ and $I_{min}$ are both proportional to the total emitted intensity and we define the degree of linear polarization of the emission as:
\begin{equation}
\delta(\Theta) = \frac{I_{max} - I_{min}}{I_{max} + I_{min}}
\end{equation} 

Including equations (\ref{eq21}) to (\ref{eq24}) in equations (\ref{eq19}) and (\ref{eq20}), one obtains, for a 1D-dipole and 2D-dipole respectively\,:

\begin{equation}
\delta(\Theta)_{1D} = \frac{C\sin^2\Theta}{(2A-2B+C)\sin^2\Theta + 2B}
\label{eq25}
\end{equation}
\begin{equation}
\delta(\Theta)_{2D} = \frac{C\sin^2\Theta}{-(2A-2B+C)\sin^2\Theta+4A+2C}
\label{eq26}
\end{equation}

In the case of a vertical dipole ($\Theta = 0$), the emission is fully unpolarized ($\delta = 0$) for both 1D and 2D dipoles, as expected given  the cylindrical symmetry of the system As the angle $\Theta$ is increased, the emission becomes more polarized and, for $\Theta = \pi/2$, $\delta$ reaches a maximum value of $C/(2A+C)$ for a 1D-dipole and $C/(2A+2B+C)$ for a 2D-dipole, which is always smaller than unity\,: the emission is never strictly fully polarized.

These equations show that, for both 1D and 2D dipoles, it is possible to extract the out-of-plane orientation $\Theta$ from the measured degree of polarization $\delta$. This requires the knowledge of the coefficients $A$, $B$ and $C$, which can be calculated theoretically for a given situation and depend on the sample configuration (presence of an interface) through the functions $f_s$, $f_{p1}$ and $f_{p2}$, and on the objective numerical aperture through $\theta_{jmax}$.\\

Let us discuss briefly the difference between excitation and emission polarization analysis. A typical \textit{excitation} polarization analysis set-up will include a rotating polarizer of angle $\alpha_{exc}$ on the path of the excitation beam, and measure the emitted intensity $I(\alpha_{exc})$. It has been shown in \cite{Ha99} that the orientation of the exciting electric field $\vec{E}_{exc}$, at the position of the emitter, is very close to the orientation $\vec{u}_{\alpha,\text{exc}}$ of the excitation polarizer, even when taking into account emitter positioning imperfections and a high objective numerical aperture. For this reason, for a 1D dipole, one can write\,:

\begin{equation}
\begin{split}
I(\alpha_{exc}) & \propto  |\vec{d}.\vec{E}_{exc}|^2 \\
& \propto I_0|\vec{u}_d.\vec{u}_{\alpha}|^2 = I_0\cos^2(\Phi-\alpha_{exc})\sin^2\Theta
\end{split}
\label{eqExc}
\end{equation} 

In this case, as pointed out since early work on single-molecule orientation \cite{Ha96,Macklin96}, the in-plane angle $\Phi$ can be obtained as the angle $\alpha_{exc}$ which maximizes $I(\alpha_{exc})$ but the out-of-plane angle $\Theta$ cannot be obtained because the value $I_0$ is not known. It is clear, by comparison of expressions (\ref{eq19}) and (\ref{eqExc}), that the polarization analysis in excitation and emission are two very different situations.\\

We now discuss the calculated correspondence between $\Theta$ and $\delta$ and analyse its physical meaning. 

We start with an emitter in a homogeneous medium (situation (i)). The values of $A$, $B$ and $C$ can then be calculated analytically \cite{Axelrod79} and in the limit of high numerical aperture ($\theta_{\text{max}} = \pi/2$) we find for a 1D dipole the simple expression\,:

\begin{equation}
\delta_{\text{high NA}}(\Theta) = \frac{7}{8}\sin^2\Theta 
\end{equation}

so that $\delta$ ranges from 0 to 0.875. On the other hand, for a 1D dipole in the limit of a low numerical aperture, a second-order development in $\theta_{\text{max}}$ leads to\,:

\begin{equation}
\delta_{\text{low NA}}(\Theta) = \frac{\sin^2\Theta}{(1-\frac{\theta_{\text{max}}^2}{2})\sin^2\Theta+\frac{\theta_{\text{max}}^2}{2}} 
\end{equation}

\begin{figure}[H]
\centering
\includegraphics[width=8cm]{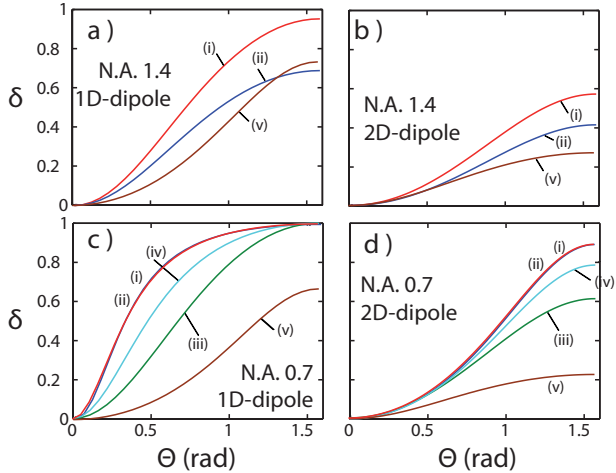}
\caption{Theoretical value of $\delta$ a function of the angle $\Theta$ for a 1D dipole (a,c) or a 2D dipole (b,d) with numerical aperture 1.4 (a,b) or 0.7 (c,d), calculated in situations (i), (ii) and (v) for (a,b) and situations (i) to (v) for (c,d) with $d$ = 50\,nm for cases (ii) and (iv) and $d$ = 0 for cases (iii) and (v), and $\lambda$ = 620\,nm.}
\label{Fig4}
\end{figure}

In this case, the maximum value of $\delta$ is 1. The case of $\theta_{max} \sim 0$ (very low numerical aperture) is interesting as it corresponds to probing a single direction of emission. Our calculations show that, for $\theta_{max}\sim 0$, $\delta$ is unity for any $\Theta$, as expected since the emission of a 1D dipole into a specific direction is always polarized (the case $\Theta = 0$ is an exception\,: it gives $\delta = 0$, but is actually not measurable because no light can be detected for this orientation\,: a 1D dipole never emits into the direction of its axis). This means that a very low numerical aperture is not an appropriate condition for measuring the orientation of an emitter by polarization analysis\,: the angle $\Theta$ cannot be deduced from the value of $\delta$ as it is always unity. As the numerical aperture is increased, the objective collects the emission into different directions, each direction having a specific polarization, so that the collected beam is a summation of different polarizations and has a lower degree of polarization. It is this summation that allows the measurement of $\Theta$ from the polarization properties. 
As for a 2D dipole in a homogeneous medium, it can also be calculated analytically that for the limit case of high numerical aperture ($\theta_{\text{max}} = \pi/2$)\,:

\begin{equation}
\delta_{\text{high NA}}(\Theta) = \frac{7}{16}\sin^2\Theta
\end{equation}

which is half the degree of polarization in the 1D case (similar trends are also observed for situations (ii) and (iv) in fig. \ref{Fig4}(a) and (b)). As expected for a 2D dipole, being a sum of two incoherent dipoles, emits with a lower degree of polarization. In the limit of low numerical aperture\,:
\begin{equation}
\delta_{\text{low NA}}(\Theta) = \frac{\sin^2\Theta}{(\frac{\theta_{\text{max}}^2}{2}-1)\sin^2\Theta+2} 
\end{equation}

For $\theta_{\text{max}}\sim 0$, which corresponds to probing a single direction of emission, $\delta$ is not unity, except for the case $\Theta$=90$^\circ$, unlike the case of a 1D dipole. This is explained by the fact that, even when probing a single direction, the emission is not necessarily polarized as it is a sum of the emissions of two incoherent dipoles with different orientations.

We now turn to the cases where the emitter is near an interface. We have introduced in section 1 several standard experimental conditions. We plot in fig. \ref{Fig4} the relation between $\Theta$ and $\delta$ for the following parameters\,: substrate index =  $1.5$, other medium index = $1$, $d$ = 50\,nm, $\lambda$ = 620\,nm. We consider the case of a 0.7 numerical aperture for situations (i) to (v), and the case of an immersion objective with 1.4 numerical aperture for situations (i), (ii) and (v). We distinguish between 1D and 2D dipoles. The obtained curves show a similar trend, with an increase from $\delta(0) = 0$ to a maximum $\delta(\pi/2)$ which is always below 1. However, the quantitative differences between these curves are significant. For a given value of $\delta$, depending on the experimental configuration and on the 1D or 2D nature of the dipole, the corresponding values of $\Theta$ can be different by up to 40$^{\text{o}}$. It is thus possible, in all these experimental configurations, to extract $\Theta$ from $\delta$, but only if the specificities of this configuration (numerical aperture, 1D/2D nature, index, interface etc) are properly included in the theoretical analysis.

In fig. \ref{Fig4}, we note that the $\Theta$-dependence of $\delta$ is similar for cases (ii) and (v)  (especially for NA=$1,4$) and for cases (iii) and (iv). This remark suggests that, for close experimental conditions (small distance $d$ between the emitter and the interface), the expected behaviour $\delta=f(\theta)$ is the same whether the emitter is slightly above or below the surface.

Finally, let us discuss the polarization analysis in the case of nanorod emission \cite{Chen01, Hu01, Rothenberg04, Pisanello10}. In these studies, the nanorods are assumed horizontally deposited ($\Theta = \pi/2$), and the degree of linear polarization $\delta$ is measured in order to probe to what extent the rod behaves as a linear dipole. Our curves show that, if the rod is a perfect 1D dipole, a value $\delta \simeq 1$ should be measured in all situations for a 0.7 numerical aperture, but, for a 1.4 numerical aperture, values of $\delta$ between 0.7 and 0.98 are calculated, depending on the situation. This must be taken into account when interpreting the experimental values of $\delta$, which range from 0.7 to 0.9 \cite{Chen01, Hu01, Pisanello10}. 


\section{Experimental orientation measurement}

In this section, we apply these considerations to demonstrate experimental orientation measurements on CdSe/CdS nanocrystals, chosen for their brightness and photostability.

\begin{figure}[H]
\centering
\includegraphics[width=8cm]
{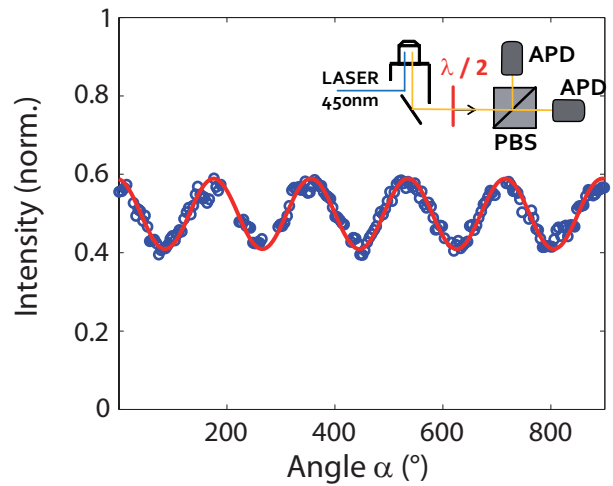}
\caption{(a) Circles\,: dependants of detected intensity as a function of polarization analysis angle for a nanocrystal of CdSe/CdS at the vicinity of air/dielectric interface. This curve is normalized by the total emitted intensity in order to account for fluctuations of the total emitted intensity. The different detection efficiencies of the two paths are corrected so that the normalized curve has a mean value of 0.5. The fitted curve (solid red line) corresponds to equation (\ref{eq20}). We deduce from the fit $\Theta$=$44^\circ$ and $\Phi$=$52^\circ$.}
\label{Fig6}
\end{figure}

We use an inverted microscope to image a sample of CdSe/CdS nanocrystals on a glass substrate, covered by 50\,nm of PMMA (measured by profilometer), with an oil-immersion objective (1.4 numerical aperture, x100). A single nanocrystal is excited by a diode laser at 450\,nm (for each measurement a standard  Hanbury-Brown and Twiss measurement demonstrates the emission of a single photon and therefore the imaging of a single emitter \cite{Brokmann04}). The photoluminescence is collected by the same objective and focused on a 100-$\mu$m pinhole in order to spatially filter the background noise. It is then recollimated, passed through a half-wave plate, separated into two arms by a polarizing beam splitter cube and each arm is focused on a single-photon counting avalanche photodiode (see inset of fig. \ref{Fig6}). The half-wave plate is continuously rotated with an angle $\alpha/2$ and the photoluminescence intensity is measured on each photodiode. The role of the half-wave plate and polarizing cube is equivalent to a polarizer of angle $\alpha$. 

We plot in fig. \ref{Fig6} the intensity on one photodiode, normalized by the sum of the intensities on the two photodiodes in order to cancel the fluctuations of the total emitted intensity, due to slight emitter instabilities. This curve is well fitted by equation (\ref{eq20}) with 3 fitting parameters\,: $I_{max}$, $I_{min}-I_{max}$ and $\Phi$. We find, for this nanocrystal, an in-plane angle $\Phi$ = 50$^{\text{o}}$ and a degree of polarization $\delta =$18\% from which we deduce, given the theoretical curve of fig. \ref{Fig4}(b) (situation (ii)), an out-of-plane angle $\Theta$ = 43$^{\text{o}}$. We estimate the precision of our fit to $\pm$ 4 $^\text{o}$ for $\Phi$, and $\pm$ 2 $^{\text{o}}$ on $\Theta$.

\begin{figure}[H]
\centering
\includegraphics[width=5cm]{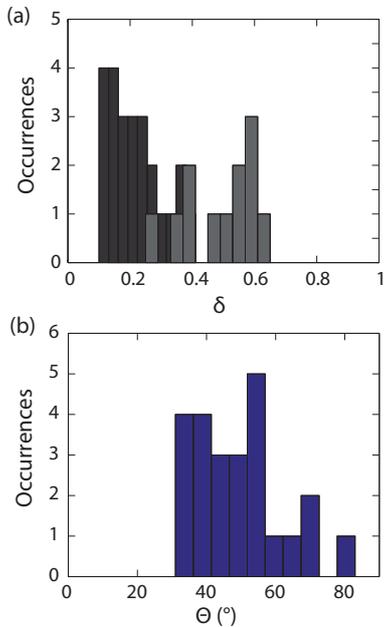}
\caption{(a) Histogram of experimental values of $\delta$ for 24  CdSe/CdS nanocrystals (blue bars) and 13 CdSe/ZnS nanocrystals (red bars). (b) Histogram of experimental values of $\Theta$ for 24 CdSe/CdS nanocrystals.}
\label{Fig7bis}
\end{figure}

We repeat this measurement for a collection of the CdSe/CdS nanocrystals and plot in fig.\ref{Fig7bis}(a) a histogram of the obtained values of $\delta$. We find values of $\delta$ below 0.4, in agreement with our theoretical calculation (fig. \ref{Fig4}(b), situation (ii)) that the value of $\delta$ ranges between 0 and 0.4 for a 2D dipole in this configuration. This is consistent with our previous demonstration that the CdSe/CdS nanocrystals are 2D emitters. On the other hand, we perform the same measurements for a collection of CdSe/ZnS nanocrystals (fig. \ref{Fig7bis}(a)) and we find values of $\delta$ up to 0.7. These values are too high for a 2D dipole, as shown by fig. \ref{Fig4}(b) (for the CdSe/ZnS nanocrystals of emission wavelength $\lambda$ = 565\,nm, the calculated curve is not shown here but very close to the case $\lambda$ = 620\,nm). They could be explained by a mixture of 1D and 2D dipoles as already proposed in section 2, since the maximum theoretical $\delta$ is 0.7 for a 1D dipole. 

We plot in fig. \ref{Fig7bis}(b) a histogram of the out-of-plane angles $\Theta$ measured for the CdSe/CdS nanocrystals. We find a distribution of angles between 31 and 83 $^{\text{o}}$. We did not find any nanocrystal with orientation $\Theta$ below 30$^{\text{o}}$. Indeed, small values of $\Theta$ are theoretically less likely\,: for an isotropic distribution of orientations $(\Theta,\Phi)$, only 13 \% of orientations show $\Theta <$ 30$^\circ$. It is also possible that the predominance of the 40-60$^\circ$ orientations is due to the specific geometry of the CdSe/CdS nanocrystals. These thick-shell NCs are not spherical and display a faceted geometry (Fig. \ref{Fig3}(d)). They tend to have a bi-pyramidal shape and they may lay on the substrate with some preferred orientation which could explain the trend we observed in the polarization measurements.

\section{Vicinity of a gold surface}

\begin{figure}[H]
\begin{center}
\includegraphics[width=8cm]{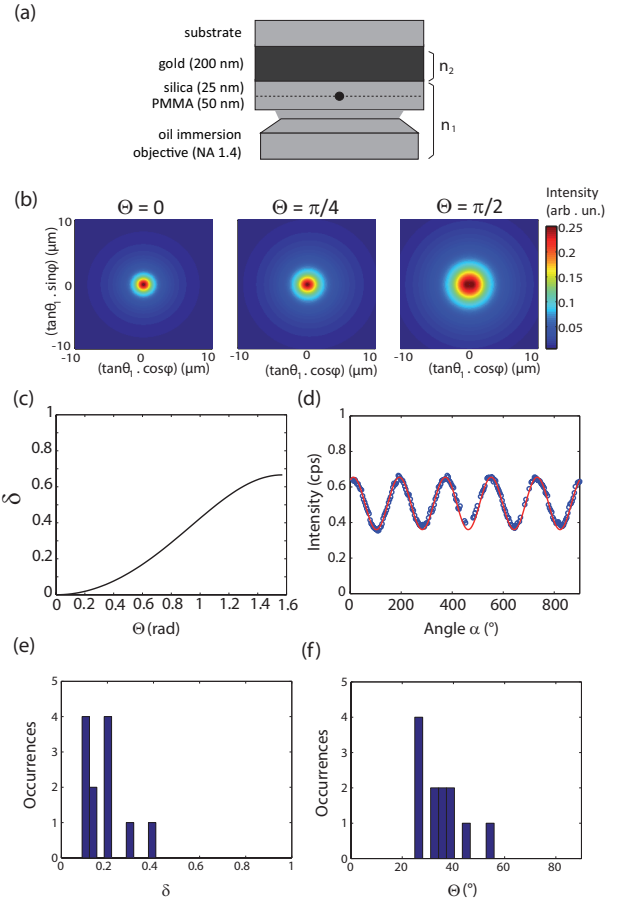}
\end{center}
\caption{(a) Schematic of the experimental system. (b) Calculated emission pattern for a 2D-dipole with ($\Theta$=$0$,$\Phi$=$0$), ($\Theta$=$\frac{\pi}{4}$,$\Phi$=$0$) and ($\Theta$=$\frac{\pi}{2}$,$\Phi$=$0$). (c) Calculated values of $\delta$ as a function of the angle $\Theta$ for a 2D-dipole (c-axis inclination). (d) Circles\,: dependence of the x-polarized emission intensity as a function of half-wave plate angle for a nanocrystal of CdSe/CdS at the vicinity of gold/dielectric interface, fitted (solid red line) with the equation (\ref{eq20}). We deduce from the fit $\Theta$=$46^\circ$ and $\Phi$=$69^\circ$. (e) and (f) Histograms of experimental values of $\delta$ and $\Theta$ for 12 CdSe/CdS nanocrystals situated at 50nm from a gold surface.}
\label{Fig8}
\end{figure}

Finally, we discuss in this last section the orientation measurement for an emitter in the vicinity of a gold film. This situation is of interest, for instance, in the context of coupling to surface plasmons \cite{Vion09} or nano antennas \cite{Belacel12}, for which the orientation is crucial. We deposit on a glass substrate 200\,nm of gold, 25\,nm of silica, CdSe/CdS nanocrystals, and 50\,nm of PMMA (fig. \ref{Fig8}(a)). We image this sample with an oil-immersion objective of numerical aperture 1.4. 

Since the optical skin depth in gold is a few tens of nanometres, the 200-nm gold layer can be considered infinitely thick. We are then in the case that we labelled as situation (ii), where $n_1 = 1.5$ is the silica/PMMA index, $d$ = 25\,nm is the distance to the gold film, $n_2$ is the gold dielectric constant (obtained from ellipsometric measurements) and $\lambda$ = 620\,nm. We plot in fig. \ref{Fig8}(b) the emission pattern, for three different angles $\Theta$. For clarity, we plot it in the 2D angular coordinates $\theta_1$ and $\phi$ expressed into cartesian coordinates ($\tan\theta_1\times\cos\phi$) and ($\tan\theta_1\times\sin\phi$). Apart from the total emitted intensity (which is not a useful quantity to deduce $\Theta$ as the emitted intensity can vary significantly among nanocrystals), there are very little differences between these images, so that $\Theta$ cannot be obtained. Moreover, the emission patterns have rotation invariance, so that $\Phi$ cannot be known from these patterns either. This is an example of a situation for which defocused imaging, which probes the emission pattern of a dipole, cannot be used to measure the orientation of a dipole.

Polarization analysis, on the other hand, is appropriate in this configuration. We plot in fig. \ref{Fig8}(c) the theoretical dependence of $\delta$ on $\Theta$. This curve shows a sufficiently clear dependence for $\Theta$ to be determined if $\delta$ is known. We plot in fig. \ref{Fig8}(d) the experimental polarization analysis for a CdSe/CdS nanocrystal. By fitting these data with eq. (\ref{eq20}), we find the nanocrystal orientation\,: $\Phi$ = 69$\pm$3$^{\text{o}}$ and $\Theta$ = 46$\pm$1$^{\text{o}}$.

We make the same measurements for 12 nanocrystals on this gold-silica substrate, and plot a histogram of the measured values of $\delta$ in fig. \ref{Fig8}(e) and of the corresponding $\Theta$ in fig. \ref{Fig8}(f). The values of $\Theta$ are distributed between 0.1 and 0.4, in agreement with the calculations of fig. \ref{Fig8}(c) showing that $\delta$ can range between 0 and 0.7. The corresponding orientations $\Theta$ range between 25 and 55$^{\circ}$. These angles are consistent with the values obtained in section 4.

\section*{Conclusion}

In this paper, we addressed the orientation measurement of a single photoluminescent emitter. We showed that, in contrast with excitation polarization analysis, emission polarization analysis provides both the in-plane angle $\Phi$ and the out-of-plane angle $\Theta$. In the case of a 2D-dipole near a gold film, polarization analysis is the most appropriate method, as the more well-established defocused imaging cannot yield precise results. We developed a model of the polarization analysis experiment, and insisted on the importance of taking into account the sample geometry (presence of an interface) and the objective numerical aperture. We distinguished five different sample configurations which we believe will cover most experimental conditions, but other situations could be easily extrapolated. We showed that the angle $\Theta$ can be deduced from the measured degree of polarization $\delta$, which is established analytically. These expressions can lead to rather different quantitative values, depending on the experimental situation and on the nature of the dipole, which have to be carefully taken into account in the model. We demonstrated experimentally the orientation of CdSe/CdS nanocrystals, which we proved to be 2D dipoles. The method we have developed here is crucial for coupling emitters to photonic and especially plasmonic systems.

\section*{Aknowledgements}
The authors would like to thank Francis Breton, St\'ephane Chenot and Lo\"ic Becerra for technical support and fabrication of the samples, as well as Jean-Louis Fave, Michel Schott and Paul B\'enalloul for fruitful discussions. This work was funded by the Agence Nationale de la Recherche (P3N Delight) and by the Centre de Comp\'etence NanoSciences Ile-de-France (C'Nano IdF).

\section*{Appendix}

In the particular case (v), one must add to the equations of case (iv) the evanescent component of the dipoles's emission to the far-field component (eq.(\ref{eq1})). This new component involves the $\vec{k_1}$ vectors which fulfil the  condition $k_1 \leq k_{\parallel_1}\leq k_2 $, where $k_1$ and $k_2$ stand for the wavevectors in the media 1 and 2 respectively. 
In this particular case, both equations (\ref{eq1}) and (\ref{eq3}) must be modified, in order to describe the added component, by replacing the unit vectors $\vec{u}_k$ by $\vec{u}_k^\ast$ and $\vec{u}_p$ by $\vec{u}_p^\ast$, defined as\,:

\begin{equation}
\vec{u}_k^\ast = \frac{1}{k_1}\left(\begin{matrix}k_{\parallel_1} \cos\phi\\ k_{\parallel_1} \sin\phi \\ i\kappa \end{matrix}\right) = 
\left(\begin{matrix}\cosh\alpha \cos\phi\\ \cosh\alpha \sin\phi \\ i\sinh\alpha \end{matrix}\right)
\end{equation}
and
\begin{equation}
\vec{u}_p^\ast = \left(\begin{matrix} i\sinh\beta \cos\phi\\ i\sinh\beta\sin\phi \\ -\cosh\beta \end{matrix}\right)
\end{equation}

with $\kappa$ the real part of the z-component of the wavevector $\vec{k_1}$ and $\beta$ the angle which verify\,:

\begin{equation}
k_{\parallel_1}^2-\kappa^2=k_1^2
\end{equation}
and
\begin{equation}
\cosh\beta=\frac{k_{\parallel_1}}{k_1} \quad \text{and} \quad \sinh\beta=\frac{\kappa}{k_1}
\end{equation}

In the same way, in order to verify (\ref{eq3}), $E_{p_a}(\theta_1,\phi)$ and $E_{p_b}(\theta_1,\phi)$, defined in (\ref{EqEpa}) and (\ref{EqEpb}), have to be changed into\,:

\begin{align}
E^\ast_{p_a}(\beta,\phi) = - \cos\Theta\cosh\beta
\end{align}
and
\begin{align}
E^\ast_{p_b}(\beta,\phi) = i\sin\Theta\sinh\beta\cos(\Phi - \phi)
\end{align}

The expression of the s-component of the electric field is expressed as\,:
\begin{align}
E^\ast_{s}(\beta,\phi) = \sin\Theta\sin(\phi - \Phi)
\end{align}
just as in equation (\ref{EqEs}).\\

One must introduce as above the functions $f_s^{\ast}$, $f_{p_a}^{\ast}$ and $f_{p_b}^{\ast}$, whose definitions for the case (v) are\,:\\

\begin{center}
\begin{tabular}{|c|c|c|c|}
\hline
   & $f_s^{\ast} =$ & $f_{p_a}^{\ast} =$ & $f_{p_b}^{\ast} =$  \\ \hline 
(v) & $t_s^{\ast 12}e^{-\kappa d}$ & $t_p^{\ast 12}e^{-\kappa d}$ & $t_p^{\ast 12}e^{-\kappa d}$ \\
\hline
\end{tabular}\end{center}

with\,:

\begin{equation}
t_s^{\ast12} = \frac{2in_1\sinh\beta}{n_2\cos\theta_2 + in_1\sinh\beta}
\end{equation}

\begin{equation}
t_p^{\ast12} = \frac{2in_1\sinh\beta}{n_1\cos\theta_2 + in_2\sinh\beta}
\end{equation}

with 
\begin{equation}
n_1 \cosh\beta = n_2 \sin\theta_2
\end{equation}

By taking into account the apodization factor $(n_1/n_2)|\cos\theta_2/\sinh\beta|^2$ for the transmission case (v) the emitted power is finally expressed as\,:

\begin{equation}
\!P^\ast(\beta) = \!\int_{\theta_2 = 0}^{\theta_{2max}} \!\int_{\phi = 0}^{2\pi} \!|{\vec{\mathcal{E}^\ast}(\beta,\phi)}.\vec{u}_{\beta}|^2 \frac{n_1}{n_2}\! \left|\frac{\cos\theta_2}{\sinh\beta}\right|^2\!\sin\theta_2d\theta_2d\phi
\end{equation}

with\,:
\begin{equation}
\begin{split}
\vec{\mathcal{E}^\ast}(\beta,\phi) = & D \frac{n_2}{n1} (E_s^{\ast}(\beta,\phi)f_s^{\ast}(\alpha)\vec{v}_s \\ & + [E^\ast_{p_a}(\beta,\phi)f_{p_a}^{\ast}(\beta)+E^\ast_{p_b}(\beta,\phi)f_{p_b}^{\ast}(\beta)]\vec{v}_p)
\end{split}
\end{equation}

Finally, for the particular case (v), the total detected power is then $\!P^\ast(\beta)$ added to $P^{(iv)}(\beta)$, the power calculated for the case (iv) in (\ref{power})\,:
\begin{equation}
\!P^{(v)}(\beta) = P^{(iv)}(\beta) + \!P^\ast(\beta)
\end{equation}

In this case, equations (\ref{eqA}), (\ref{eqB}) and (\ref{eqC}) must be modified as well, to the following\,:

\begin{equation}
A^\ast = \int_{\theta_2=\asin\frac{n_1}{n_2}}^{\pi/2} D^2\frac{\pi}{4} \frac{n_2}{n_1} \left|\frac{\cos\theta_{2}}{\sinh\beta}\right|^2 |f_s^\ast - i\sinh\beta f_{p_b}^\ast|^2\sin\theta_2 d\theta_2
\end{equation}
\begin{equation}
B^\ast = \int_{\theta_2=\asin\frac{n_1}{n_2}}^{\pi/2} D^2\pi  \frac{n_2}{n_1}\left|\frac{\cos\theta_{2}}{\sinh\beta}\right|^2 |f_{p_a}^\ast\cosh\beta|^2\sin\theta_2 d\theta_2
\end{equation}
\begin{equation}
C^\ast = \int_{\theta_2=\asin\frac{n_1}{n_2}}^{\pi/2} D^2\frac{\pi}{2} \frac{n_2}{n_1} \left|\frac{\cos\theta_{2}}{\sinh\beta}\right|^2 |i\sinh\beta f_{p_b}^\ast+f_s^\ast|^2\sin\theta_2 d\theta_2
\end{equation}

\end{multicols}

\end{document}